\documentclass[twocolumn,aps,pra,superscriptaddress,showpacs,tightenlines]{revtex4-1}
\usepackage[T1]{fontenc}
\usepackage[latin9]{inputenc}
\usepackage{amsmath}
\usepackage{graphicx}
\usepackage{amssymb}

\begin{document}

\title{Protecting Dissipative Quantum State Preparation via Dynamical Decoupling}

\author{Z. R. Gong}
\author{Wang Yao}

\affiliation{Department of Physics, and Center for Theoretical and Computational Physics, The University of Hong Kong, China}

\begin{abstract}
We show that dissipative quantum state preparation processes can be protected against qubit dephasing by interlacing the state preparation control with dynamical decoupling (DD) control consisting of a sequence of short $\pi$-pulses. The inhomogeneous broadening can be suppressed to second order of the pulse interval, and the protection efficiency is nearly independent of the pulse sequence but determined by the average interval between pulses. The DD protection is numerically tested and found to be efficient against inhomogeneous dephasing on two exemplary dissipative state preparation schemes that use collective pumping to realize many-body singlets and linear cluster states respectively. Numerical simulation also shows that the state preparation can be efficiently protected by $\pi$-pulses with completely random arrival time. Our results make possible the application of these state preparation schemes in inhomogeneously broadened systems. DD protection of state preparation against dynamical noises is also discussed using the example of Gaussian noise with a semiclasscial description.
\end{abstract}
\pacs{03.67.Bg, 03.67.Pp, 42.50.Dv, 03.65.Yz}

\maketitle
\section{Introduction}

Dissipative quantum state preparation has recently emerged as a conceptually
new approach for realizing resources of multipartite quantum entanglement.
In the control of an open quantum system, dissipative channels are
usually the sources of errors and play deleterious
roles. However, by properly tailoring the dissipative channels in
an open quantum system, the irreversible dynamics can have steady
states containing the important resource of quantum entanglement. Earlier studies in small scale
systems have shown that two-qubit entanglement can be generated in
the steady states of various dissipative dynamics driven either by
external incoherent pumping or by the tailored coupling with the environment~\cite{1,2,3,4}.
Recent studies have lead to the discovery of various dissipative
state preparation approaches for realizing multipartite entanglement
in large scale systems~\cite{5,6,7,8,9,10,11,12,13}. In particular,
it has been shown that by using only a few incoherent pumping channels
that couple collectively to the qubits, various types of multipartite
entanglement can be prepared, including the linear cluster states~\cite{10},
many-body singlet states~\cite{11}, and symmetric and asymmetric
Dicke states~\cite{12,13}. These multipartite entangled states can be the crucial resource for measurement based quantum information processing~\cite{1b,2b,3b,4b}. With the desired quantum states unconditionally
realizable as steady states of the irreversible dynamics, the dissipative state preparation schemes have the advantage of robustness and do not need accurate temporal controls as compared to conventional state preparation by coherent evolutions.

Most dissipative state preparation schemes are proposed for qubit systems with homogeneous resonances~\cite{1,2,3,4,10,11,12,13}.
In reality, the qubit resonance can have inevitable inhomogeneous broadening which leads to inhomogeneous dephasing. Moreover, fluctuation of the environment can also induce dynamical noise that leads to dephasing as well. These unwanted noise channels will compete with the tailored
dissipative channels for preparing quantum entanglement in the irreversible dynamics. The
fidelity of the steady state with the target quantum state will decrease
with the increase of the noise strength and the number of qubits.
These inevitable noises will render the state preparation scheme
impractical in large scale systems in the presence of inhomogeneous broadening and dynamical noise. Whether or not we can suppress the noises while preserving the desired dissipative
channels in the irreversible dynamics becomes the key.

Dynamical decoupling (DD) has been an extensively explored approach
in high-precision magnetic resonance spectroscopy~\cite{14,15,16,17},
and for suppressing the effect of qubit dephasing in quantum information processing~\cite{18,19,20,21}.
The basic idea is to introduce a sequence of pulses on the qubit in
order to average out the inhomogeneous broadening and the unwanted coupling with the environment,
hence eliminating the noise effects on the qubit. Past studies on
dynamical decoupling have focused on how to freeze the evolution of a qubit prepared on an arbitrary quantum state such that the quantum memory time can be enhanced. One efficient DD approach for protecting quantum
memory uses sequences of instantaneous $\pi$-pulses with properly
chosen arrival time. Inhomogeneous dephasing can be
removed transiently at a certain time known as the spin echo time, while
dynamical noises can also be suppressed to high orders at the spin
echo times depending on the pulse sequences. The protection of coherent evolutions by DD control pulses has also been studied~\cite{36,gate1,gate,gate2,gate3}.

In this paper, we show that dissipative quantum state preparation processes can be protected against qubit dephasing by interlacing the dissipative control with DD control consisting of sequences of $\pi$-pulses.
The DD control is introduced here to suppress the qubit dephasing while preserving a nontrivial irreversible evolution for generating entanglement in the steady state.
For suppression of inhomogeneous dephasing, we compare DD controls consisting of repetitions of a basic unit which takes various designs including the Carr-Purcell-Meiboom-Gill (CPMG)~\cite{22,23}, concatenated (CDD)~\cite{24,25,26,27,28,29}, and Uhrig sequences (UDD)~\cite{30,31,32,33}. We found the leading noise term in the Magnus expansion is of second order of the pulse interval and the residue effect of inhomogeneous dephasing is largely determined by the average pulse interval only, nearly independent of the types of the DD pulse sequences. High fidelity state preparation is therefore possible even in the presence of substantial inhomogeneous broadening when the DD pulse sequence with sufficiently small pulse interval is applied. DD protection from dynamical noise is discussed for the example of Gaussian noise of a semiclassical description, where the leading noise effect is also found to be of second order of the pulse interval. The order of noise suppression is qualitatively different from that in the DD protection of quantum memory because of the presence of the state-preparation channel, and its interference with the noise channels determines the residue noise effect.

For the scheme of preparing the many-body singlet~\cite{11,12},
we perform systematic numerical simulations of the DD protection against inhomogeneous dephasing using
CPMG, concatenated, and Uhrig pulse sequences. We find that the Magnus expansion converges well and the residue inhomogeneous broadening effect
is well accounted by the leading noise term in the Magnus series. Numerical simulation confirms that the
protection efficiency is nearly independent of the pulse sequence and depends only
on the average pulse interval. Moreover, the application of $\pi$-pulses
with completely random arrival time~\cite{34,35} is found to be efficient
in suppressing the effect of inhomogeneous broadening. Thus, the DD protection is robust against errors in the pulse arrival time. We also explain how to implement the DD protection on the scheme
of preparing linear cluster states~\cite{10}, where our numerical simulation confirms that high fidelity state preparation can be realized in the presence of substantial inhomogeneous broadening through the DD control. These dissipative preparation schemes for preparing the two important classes of multipartite entanglement are therefore applicable to inhomogeneously broadened qubit systems.

The paper is organized as follows. In section II, we discuss
the basic idea of protecting dissipative quantum state preparation
against noise channels by interlacing the state-preparation control with the DD control pulses. The evolution of the system is expanded in terms
of the Magnus series. In section III, we discuss the DD protection of the state preparation against inhomogeneous
dephasing, and derive coefficients of Magnus series for investigating
the order of noise suppression. In section IV, we discuss the DD protection on the preparation of many-body singlet
states against inhomogeneous dephasing and show numerical simulation results. In section V, we numerically study the DD protection of  the linear cluster states in the presence of inhomogeneous dephasing. In section VI, DD protection against dynamical noise is discussed for the example of Gaussian noise of a semiclassical description. Finally, the conclusions are summarized in section
VII.

\section{Magnus Expansion}

In the dissipative quantum state preparation, the evolution of the
open quantum system is described by the master equation as $\dot{\rho}=\mathcal{L}_{\mathrm{P}}\left[\rho\right]$,
where $\rho$ is the density matrix of the system and $\mathcal{L}_{\mathrm{P}}\left[\rho\right]$
is the super-operator describing the dissipative dynamics tailored
for generating entanglement. In a typical dissipative quantum state
preparation process~\cite{1,2,3,4,5,6,7,8,9,10,11,12,13}, the
dissipative channels $\mathcal{L}_{\mathrm{P}}$ are time independent
and are engineered such that the desired entangled states are obtained
as the steady states of the irreversible dynamics.

In reality, there exist other noise channels which also affect the
system dynamics as denoted by the super-operator $\mathcal{L}_{\mathrm{N}}\left[\rho\right]$,
which can have various origins such as inhomogeneous broadening of
the level spacing, and dynamical fluctuation from the coupling with an environment. The unwanted evolution $\mathcal{L}_{\mathrm{N}}\left[\rho\right]$
will compete with the tailored dissipative dynamics $\mathcal{L}_{\mathrm{P}}\left[\rho\right]$.
As a result, the steady state of the dynamics $\dot{\rho}=\mathcal{L}_{\mathrm{P}}\left[\rho\right]+\mathcal{L}_{\mathrm{N}}\left[\rho\right]$
deviates from the target entangled state. The state preparation fidelity
will drop with the increase of the noise strength and the number of
qubits. To achieve high fidelity state preparation in large scale
systems, the undesired noise channels $\mathcal{L}_{\mathrm{N}}$
need to be suppressed while preserving the desired dissipative dynamics
$\mathcal{L}_{\mathrm{P}}\left[\rho\right]$ at the same time.

\begin{figure}[ptb]
\includegraphics[bb=1 264 552 619,clip,width=3in]{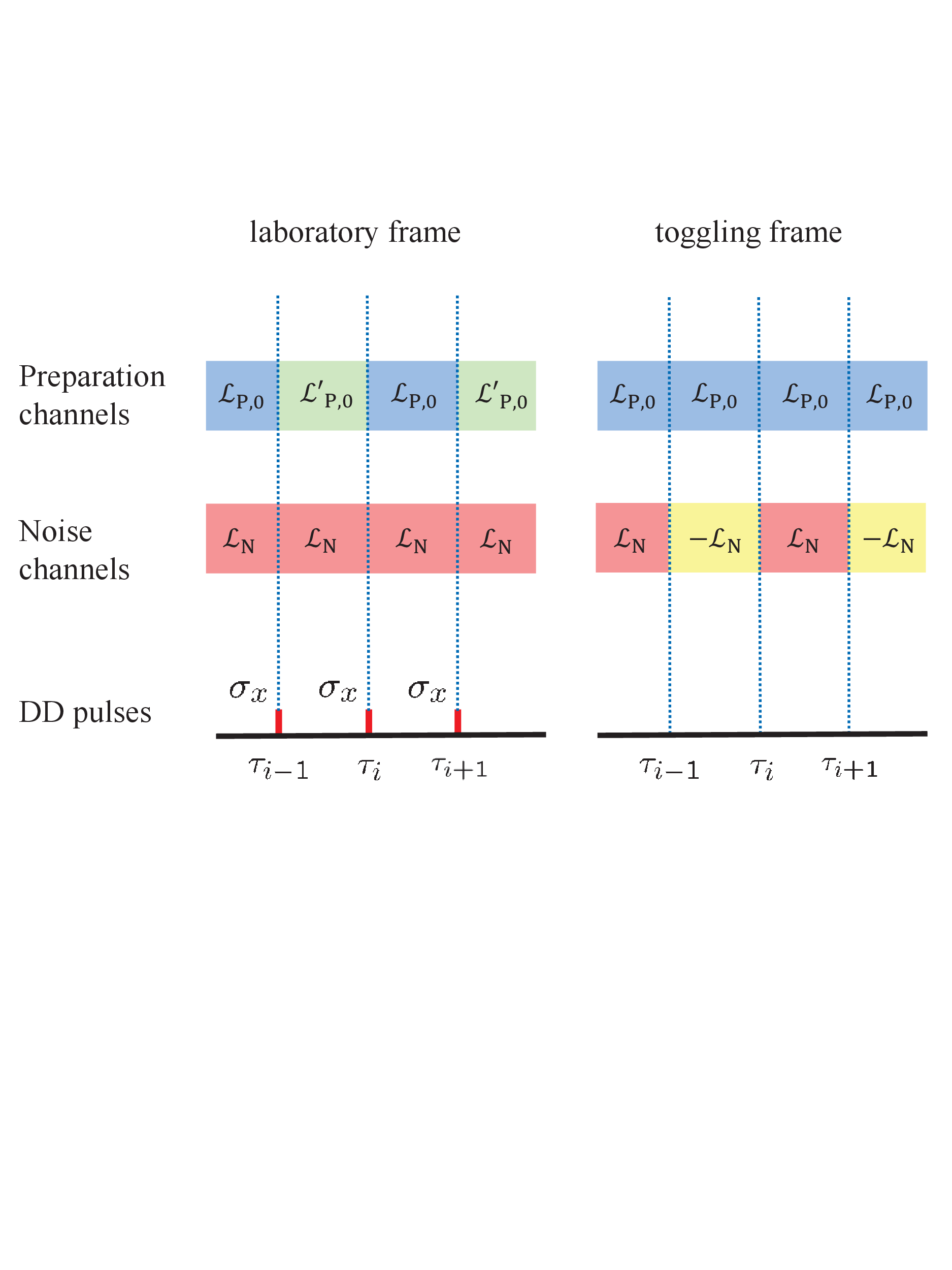}
\caption{(Color online) Schematics of interlacing DD controls with the dissipative
quantum state preparations. In the toggling frame, the noise has a
sign change between odd and even time intervals and the noise effects
average out over time. The effects of the dissipative preparation channel
in the odd and even time intervals constructively add. In the laboratory
frame, the dissipative preparation channel shall then take two different
forms for the odd and even time intervals in general.}
\label{fig:fig1}
\end{figure}

We consider interlacing the state-preparation
control with the DD control consisting of a sequence of short pulses which realize the unitary rotations
$U_{i}$ at time $\tau_{i}$ on all qubits. If the duration of the
pulse is much smaller than the pulse interval, the rotation can be
considered as instantaneous. It is convenient to analyze the evolution
of the system in the toggling frame that follows these rotations of
the qubits. In the toggling frame, the super-operators for the noise
channels and the desired dissipative channels become: $\mathcal{L}_{\mathrm{N}}^{\mathrm{T}}\left[t;\rho\right]=U(t)\mathcal{L}_{\mathrm{N}}\left[\rho\right]U(t)^{\dagger}$
and $\mathcal{L}_{\mathrm{P}}^{\mathrm{T}}\left[t;\rho\right]=U(t)\mathcal{L}_{\mathrm{P}}\left[\rho\right]U(t)^{\dagger},$
where $U(t)\equiv\prod_{j=1}^{i}U_{j}$ for $\tau_{i}<t<\tau_{i+1}$.
Below we focus on DD control using sequences of short $\pi$ pulses
(see Fig.~\ref{fig:fig1}). In such case, $U_{i}=\sigma_{x}$, and
we have $U(t)=\sigma_{x}$ in the odd intervals (i.e. $\tau_{i}<t<\tau_{i+1}$
with odd $i$), and $U(t)=I$ in the even intervals (i.e. $\tau_{i}<t<\tau_{i+1}$
with even $i$).

If the noise super-operators in the toggling frame have overall
sign changes for adjacent intervals as $\mathcal{L}_{\mathrm{N}}^{\mathrm{T}}\left[t;\rho\right]=f(t)\mathcal{L}_{\mathrm{N}}\left[\rho\right]$,
where $f(t)=(-1)^{i}$ for $\tau_{i}<t<\tau_{i+1}$, the effect of
the noise can then be averaged out over time. This is the case for pure dephasing noise. However, in the protection
of dissipative quantum state preparation, the desired dissipative
channels for generating entanglement shall also be preserved. This
imposes an additional requirement~\cite{36,gate}. The simplest scenario is to require the dissipative
channel to always take the desired form $\mathcal{L}_{\mathrm{P},0}$ in the toggling frame, i.e.
$\mathcal{L}_{\mathrm{P}}^{\mathrm{T}}\left[t;\rho\right]=\mathcal{L}_{\mathrm{P},0}\left[\rho\right]$. Thus, in the laboratory frame, the super-operator for the dissipative channel shall alternate between the two forms $\mathcal{L}_{\mathrm{P},0}\left[\rho\right]$
and $\mathcal{L}'_{\mathrm{P},0}\left[\rho\right]\equiv\sigma_{x}\mathcal{L}_{\mathrm{P},0}\left[\rho\right]\sigma_{x}$
in the even and odd intervals respectively (see Fig.~\ref{fig:fig1}). In general, $\mathcal{L}_{\mathrm{P},0}$  and $\mathcal{L}_{\mathrm{P},0}'$ are different operators. Thus, for a dissipative state preparation protocol to be protectable
by DD controls consisting of $\pi$ pulses, the dissipative dynamics
shall switch between the two forms each time a $\pi$ pulse is applied
if $\left[\sigma_{x},\mathcal{L}_{\mathrm{P}}\left[\rho\right]\right]\neq0$.
This can indeed be implemented in most dissipative preparation protocols
as the dissipative channels $\mathcal{L}_{\mathrm{P}}\left[\rho\right]$ are engineered ones.

In the toggling frame, the super-operator for the dissipative
preparation channel is time-independent, hence
the effect adds constructively over time. The super-operator for the noise channels has alternating
signs for even and odd time intervals, hence the effect destructively
cancels. The master equation in the toggling frame is: \begin{eqnarray}
\dot{\rho} & = & \mathcal{L}_{\mathrm{P},0}[\rho]+f(t)\mathcal{L}_{\mathrm{N}}\left[\rho\right],\label{eq:2-1}\end{eqnarray}
which has the formal solution: \begin{eqnarray}
\rho\left(t\right) & = & \mathcal{T}\exp\left[\int_{0}^{t}\left(\mathcal{L}_{\mathrm{P},0}+f(t')\mathcal{L}_{\mathrm{N}}\right)dt'\right]\rho\left(0\right)\label{eq:2-2}\end{eqnarray}
 with time ordering $\mathcal{T}$ and initial density matrix $\rho\left(0\right)$.
Eq. (\ref{eq:2-2}) can be rewritten in terms of the Magnus expansion:
\begin{eqnarray}
\rho\left(t\right) & = & \exp\left[\Omega\left(t\right)\right]\rho\left(0\right).\label{eq:2-3}\end{eqnarray}
Here, the formal evolution operator is expanded in terms of the Magnus
series as $\Omega\left(t\right)=\sum_{k=1}^{\infty}\Omega_{k}\left(t\right)$,
and the first three terms of this series are: \begin{subequations}
\begin{align}
\Omega_{1}\left(t\right) & =\int_{0}^{t}\mathcal{L}\left[t_{1}\right]dt_{1},\label{eq:2-4-1}\\
\Omega_{2}\left(t\right) & =\frac{1}{2}\int_{0}^{t}\int_{0}^{t_{1}}\left[\mathcal{L}\left[t_{1}\right],\mathcal{L}\left[t_{2}\right]\right]dt_{1}dt_{2},\label{eq:2-4-2}\\
\Omega_{3}\left(t\right) & =\frac{1}{6}\int_{0}^{t}\int_{0}^{t_{1}}\int_{0}^{t_{2}}\left\{ \left[\mathcal{L}\left[t_{1}\right],\left[\mathcal{L}\left[t_{2}\right],\mathcal{L}\left[t_{3}\right]\right]\right]+\right.\nonumber \\
 & \left.\left[\mathcal{L}\left[t_{3}\right],\left[\mathcal{L}\left[t_{2}\right],\mathcal{L}\left[t_{1}\right]\right]\right]\right\} dt_{1}dt_{2}dt_{3},\label{eq:2-4-3}\end{align}
 \label{eq:2-4} \end{subequations}
where $\mathcal{L}\left[t\right]=\mathcal{L}_{\mathrm{P},0}+f(t)\mathcal{L}_{\mathrm{N}}$,
and $\left[\mathcal{L}\left[t_{1}\right],\mathcal{L}\left[t_{2}\right]\right]\rho\equiv\mathcal{L}\left[t_{1};\mathcal{L}\left[t_{2};\rho\right]\right]-\mathcal{L}\left[t_{2};\mathcal{L}\left[t_{1};\rho\right]\right]$
is the Poisson bracket for the super-operator $\mathcal{L}\left[t\right].$
The Magnus expansion is convergent if $\int_{0}^{t}\left\Vert \mathcal{L}\left[t'\right]\right\Vert dt'<\pi$,
where $\left\Vert \mathcal{L}\right\Vert $ is a two-form of super-operator
$\mathcal{L}$~\cite{37}. It is convenient to investigate the suppression
of noise channels with the Magnus expansion. The residue effects of noises
are characterized by the leading term in the Magnus series that contains
the noise operators, as we will discuss explicitly in the next section
for inhomogeneous dephasing noise.

\begin{figure}[t]
\includegraphics[bb=44 386 531 663,clip,width=3in]{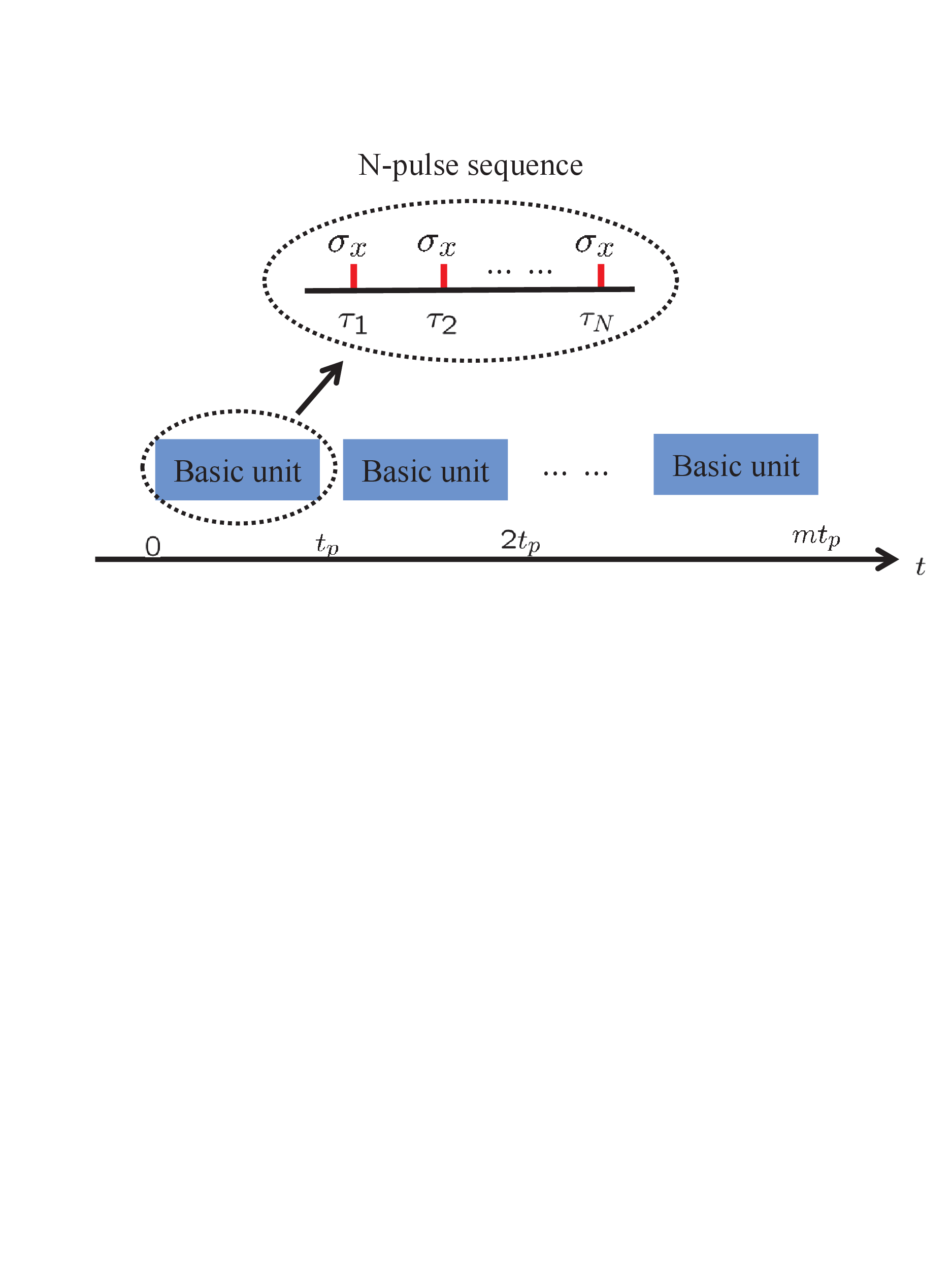}
\caption{(Color online) Schematics of DD control, formed by repetitions of a basic unit with
duration $t_{p}$. The basic unit consists of a $N$-pulse sequence
which can take various designs, including the CPMG,
concatenated and Uhrig sequences.}
\label{fig:fig2}
\end{figure}

\section{Protection of dissipative state preparation from inhomogeneous dephasing}

\label{inhomoanalysis}

In this section, we investigate the protection of dissipative state
preparation from inhomogeneous dephasing by the DD controls. The super-operator
for this noise channel is \begin{eqnarray}
\mathcal{L}_{\mathrm{N}}\left[\rho\right] & = & -i\left[\sum_{i}\omega_{i}\sigma_{i}^{z},\rho\right],\label{eq:3-1}\end{eqnarray}
where the inhomogeneous dephasing is the consequence of inhomogeneous
broadening in the qubit resonance $\omega_{i}$. The first three terms
of the Magnus expansion then become: \begin{subequations} \begin{eqnarray}
\Omega_{1}\left(t\right) & = & t\left[\mathcal{L}_{\mathrm{P},0}+c_{1}\left(t\right)\mathcal{L}_{\mathrm{N}}\right],\label{eq:3-2-1}\\
\Omega_{2}\left(t\right) & = & t^{2}c_{2}\left(t\right)\left[\mathcal{L}_{\mathrm{P},0},\mathcal{L}_{\mathrm{N}}\right],\label{eq:3-2-2}\\
\Omega_{3}\left(t\right) & = & t^{3}c_{3,a}\left(t\right)\left[\mathcal{L}_{\mathrm{P},0},\left[\mathcal{L}_{\mathrm{P},0},\mathcal{L}_{\mathrm{N}}\right]\right]\nonumber \\
 &  & +t^{3}c_{3,b}\left(t\right)\left[\mathcal{L}_{\mathrm{N}},\left[\mathcal{L}_{\mathrm{P},0},\mathcal{L}_{\mathrm{N}}\right]\right],\label{eq:3-2-3}\end{eqnarray}
 \label{eq:3-2} \end{subequations}
where $\left[\mathcal{L}_{\mathrm{P},0},\mathcal{L}_{\mathrm{N}}\right]\rho=\mathcal{L}_{\mathrm{P},0}\left[\mathcal{L}_{\mathrm{N}}\left[\rho\right]\right]-\mathcal{\mathcal{L}}_{\mathrm{N}}\left[\mathcal{L}_{\mathrm{P},0}\left[\rho\right]\right]$
stands for the Possion bracket for the super-operators. Because of the interference between the super-operators $\mathcal{L}_{\mathrm{P},0}$
and $\mathcal{L}_{\mathrm{N}}$, these Poisson brackets are typically
nonzero.

\begin{center}
\begin{table*}[t]
 \caption{Coefficients of leading Magnus terms in Eq.~(\ref{3-5}). The coefficients
are evaluated for one basic unit of evolution with duration $t_{p}$
where the DD control is a $N$-pulse sequence taking various designs
(see text). }
\label{tab:coefficient} \begin{tabular}{|c|c|c|c|c|c|c|c|c|}
\hline
 & no pulse  & CPMG  & CDD3  & CDD4  & UDD3  & UDD4  & UDD5\tabularnewline
\hline
\hline
$N$  & 0  & 2  & 5  & 10  & 3  & 4  & 5 \tabularnewline
\hline
$\alpha_{1}$  & 1  & 0  & 0  & 0  & 0  & 0  & 0\tabularnewline
\hline
$\alpha_{2}$  & 0  & 0  & 0  & 0  & 0  & 0  & 0\tabularnewline
\hline
$\alpha_{3,a}$  & 0 & $3.12\times10^{-2}$  & 0  & 0  & 0  & 0  & 0\tabularnewline
\hline
$\alpha_{3,b}$  & 0 & $-1.04\times10^{-2}$  & $-2.60\times10^{-3}$  & $-6.51\times10^{-4}$  & $-5.05\times10^{-3}$  & $-3.04\times10^{-3}$  & $-2.04\times10^{-3}$\tabularnewline
\hline
$\alpha_{3,b}\times N^{2}$  & 0  & $-4.16\times10^{-2}$  & $-6.51\times10^{-2}$  & $-6.51\times10^{-2}$  & $-4.55\times10^{-2}$  & $-4.86\times10^{-2}$  & $-5.11\times10^{-2}$ \tabularnewline
\hline
\end{tabular}
\end{table*}
\par\end{center}

The coefficients in Eq.~(\ref{eq:3-2}) are given by: \begin{subequations}
\begin{eqnarray}
c_{1}\left(t\right) & = & \frac{1}{t}\int_{0}^{t}f\left(t_{1}\right)dt_{1},\label{eq:3-3-1}\\
c_{2}\left(t\right) & = & \frac{1}{2t^{2}}\int_{0}^{t}t_{1}g\left(t_{1}\right)dt_{1},\label{eq:3-3-2}\\
c_{3,a}\left(t\right) & = & \frac{1}{12t^{3}}\int_{0}^{t}t_{1}^{2}h\left(t_{1}\right)dt_{1},\label{eq:3-3-3}\\
c_{3,b}\left(t\right) & = & \frac{1}{12t^{3}}\int_{0}^{t}t_{1}^{2}\left[6c_{2}\left(t_{1}\right)f\left(t_{1}\right)-c_{1}\left(t_{1}\right)g\left(t_{1}\right)\right],\notag\\
\label{eq:3-3-4}
\end{eqnarray}
\label{eq:3-3}
\end{subequations}
Here $g\left(t_{1}\right)\equiv c_{1}\left(t_{1}\right)-f\left(t_{1}\right)$
and $h\left(t_{1}\right)\equiv6c_{2}\left(t_{1}\right)-g\left(t_{1}\right)$.
Since $f(t_{1})=(-1)^{i}$ for $\tau_{i}<t_{1}<\tau_{i+1}$, if the
total duration of the odd intervals equals that of the even intervals,
$c_{1}$ will vanish and the leading term of the Magnus expansion
$\Omega_{1}$ only contains the ideal dissipative preparation channel
we wish to preserve. The residue effect of the inhomogeneous broadening
is given by the higher order Magnus terms, which results from the
interference of the noise channel with the preparation channel.

In the dissipative state preparation, the desired quantum state is
obtained as the steady state of the dynamics, and the timescale to
reach the steady state is determined by the dissipative channel $\mathcal{L}_{P,0}$.
We consider a practical scenario of interlacing the DD control with
the dissipative preparation (see Fig.~\ref{fig:fig2}). The DD control
has a periodic structure which consists of repetitions of a basic
unit with the duration $t_{p}$. The basic unit is an $N$-pulse sequence
that can take various designs, including the CPMG~\cite{22,23},
concatenated~\cite{24,25,26,27,28,29}, and Uhrig~\cite{30,31,32,33}
sequences. The basic unit of the DD control can be repeated as many
times as necessary along with the dissipative evolution until a steady
state is reached. The average interval between the $\pi$-pulses can
be defined $\bar{\tau}\equiv t_{p}/N$.

For a periodic control with period $t_{p}$, it has been proved that
$\Omega_{k}\left(lt_{p}\right)=l\Omega_{k}\left(t_{p}\right)$, where
$l=1,2,\ldots.$~\cite{37}. Therefore the coefficients $c_{n}$
possess the scaling law \begin{eqnarray}
c_{n}\left(lt_{p}\right) & = & l^{1-n}c_{n}\left(t_{p}\right)=\left(\frac{t_{p}}{t}\right)^{n-1}\alpha_{n}.\label{eq:3-4}\end{eqnarray}
Here $\alpha_{n}\equiv c_{n}(t_{p})$ depends on the structure of
the $N$-pulse unit, but is independent of its duration $t_{p}$ as
is evident from Eq.~(\ref{eq:3-3}). The Magnus terms then have the
form \begin{subequations} \begin{eqnarray}
\Omega_{1}\left(t=lt_{p}\right) & = & t\left[\mathcal{L}_{\mathrm{P},0}+\alpha_{1}\mathcal{L}_{\mathrm{N}}\right],\label{eq:3-5-1}\\
\Omega_{2}\left(t=lt_{p}\right) & = & t(\alpha_{2}t_{p})\left[\mathcal{L}_{\mathrm{P},0},\mathcal{L}_{\mathrm{N}}\right],\label{eq:3-5-2}\\
\Omega_{3}\left(t=lt_{p}\right) & = & t\left(\alpha_{3,a}t_{p}^{2}\right)\left[\mathcal{L}_{\mathrm{P},0},\left[\mathcal{L}_{\mathrm{P},0},\mathcal{L}_{\mathrm{N}}\right]\right]\nonumber \\
 &  & +t\left(\alpha_{3,b}t_{p}^{2}\right)\left[\mathcal{L}_{\mathrm{N}},\left[\mathcal{L}_{\mathrm{P},0},\mathcal{L}_{\mathrm{N}}\right]\right].\label{eq:3-5-3}\end{eqnarray}
 \label{3-5} \end{subequations}
 Obviously, for the system evolution
in a given duration $t$, the unwanted dynamics described by the higher
order Magnus terms $\Omega_{k\geq2}$ can be better suppressed when
$t_{p}$ is smaller for a given type of $N$-pulse unit. This is equivalent
to using pulse sequences with shorter interval $\bar{\tau}$. When
$t_{p}$ is sufficiently small, the residue effects of noises are
predominantly captured by the leading noise term, i.e. $\Omega_{k\geq2}$
with the smallest $k$, and convergence of the Magnus expansion can
be expected.

In table I, we list the coefficients $\alpha_{n}$ for the first several
Magnus terms, where the basic unit uses various DD pulse sequences.
In the DD protection of quantum memory, it is well known that inhomogeneous
dephasing can be completely removed, and dynamic noises can be suppressed
to arbitrarily specified order of the pulse interval by using higher
order concatenation design~\cite{24,25,26,27,28,29} or the Uhrig
design~\cite{30,31,32,33}. Here for the protection of the state
preparation, we find a qualitative different behavior. Because of
the presence of nontrivial dissipative dynamics for the state preparation,
the inhomogeneous broadening always has residue effects due to its
interference with the preparation channel (c.f. Eq.~(\ref{eq:3-2})).
For all pulse sequences considered, $\alpha_{3,b}$ can only be suppressed but never vanish,
and the inhomogeneous noise can be suppressed at most to the second
order of the pulse interval.

An interesting observation is $\alpha_{3,b}N^{2}$
is nearly independent of the type of pulse sequences. We note that
the Magnus expansion in such cases reads \begin{equation}
\Omega(t)=t\left(\mathcal{L}_{P,0}+\alpha_{3,b}N^{2}\bar{\tau}^{2}\left[\mathcal{L}_{\mathrm{N}},\left[\mathcal{L}_{\mathrm{P},0},\mathcal{L}_{\mathrm{N}}\right]\right]+O(\bar{\tau}^{3})\right)\label{leading}\end{equation}
where we have used $t_{p}=N\bar{\tau}$. Thus, the efficiency on
the suppression of inhomogeneous dephasing is mostly determined by
the average pulse density $\bar{\tau}$, instead of the order of the
CDD or UDD sequences.

Here we note that $\mathcal{L}_{P,0}$ in the above derivation can stands for a general evolution that one wishes to preserve while using DD control to suppress the dephasing.

\begin{figure}[ptb]
\begin{centering}
\includegraphics[bb=21 362 567 792,clip,width=3.5in]{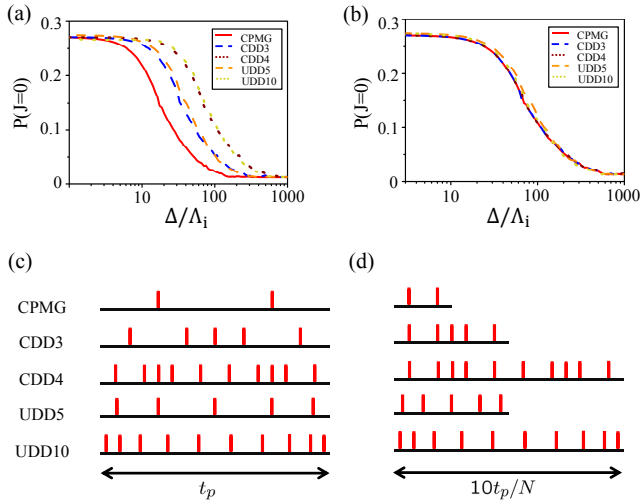}
\par\end{centering}
\caption{(Color online) Steady state population of many-body singlet $P(J=0)$
versus the strength of inhomogeneous broadening $\Delta$, when the preparation process
is protected using different DD pulse sequences. The ensemble consists
of 6 qubits, with inhomogeneously broadened energies $\omega_{i}=\Delta(7-2i)/5$,
$i=1,2,\ldots,6$. The collective pumping rates are chosen as $\Lambda_{h}=10\Lambda_{i}$
(see text). The red solid, blue dashed, brown dotted, orange dashed,
and light yellow dotted curves show numerical results by solving the exact
master equation, where the basic unit of the DD control consists of
the CPMG, CDD3, CDD4, UDD5 and UDD10 pulse sequences respectively.
For the simulation presented in (a), the duration of the basic unit
$t_{p}=10^{-3}\Lambda_{i}^{-1}$ for all DD controls compared, as
shown in (c). In (b), the duration of the basic unit $t_{p}=10^{-4}\Lambda_{i}^{-1}N$,
$N$ being the number of pulses in the basic unit, and the average
pulse interval $\bar{\tau}\equiv t_{p}/N$ is taken to be the same
for all DD controls compared, as shown in (d). }
\label{fig:fig3}
\end{figure}

\begin{figure}[ptb]
\begin{centering}
\includegraphics[bb=2 284 586 800,clip,width=3.5in]{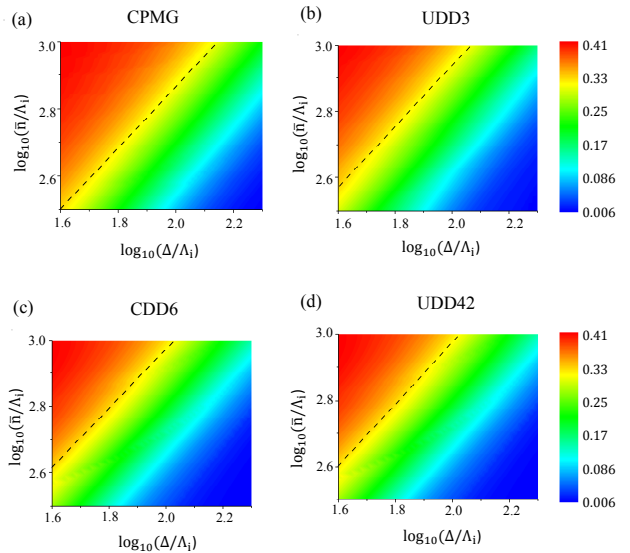}
\par\end{centering}
\caption{(Color online) Steady state population of many-body singlet $P(J=0)$
as function of the inhomogeneous broadening $\Delta$ and the average
pulse density $\bar{n}=N/t_{p}$, where the basic unit of DD control
uses various $N$-pulse sequences. The ensemble consists of 6 qubits,
with inhomogeneously broadened energies $\omega_{i}=\Delta(7-2i)/5$, $i=1,2,\ldots,6$.
The collective pumping rates are chosen as $\Lambda_{h}=100\Lambda_{i}$.}
\label{fig:fig4}
\end{figure}

\section{DD protection on the dissipative preparation of many-body singlets}

In this section, we numerically study the DD protection on an exemplary dissipative state preparation scheme in the presence of inhomogeneous dephasing. The scheme uses collective pumping to prepare many-body singlets~\cite{11}. It has been shown that when
an ensemble of spins are collectively pumped by the homogeneous collective
lowering operator $\sum_{n}\hat{s}_{n}^{-}$ and collective raising
operators $\sum_{n}a_{n}\hat{s}_{n}^{+}$ with inhomogeneous coefficients
$a_{n}$, the steady state is a highly entangled one in the neighborhood
of many-body singlets. In the absence of inhomogeneous and homogeneous
dephasing, the singlet population in the steady state $P(J=0)\geq20\%$.
Here we consider the scenario on the preparation of many-body singlet
as investigated in Ref.~\cite{12}. An spin ensemble with the even number
of spins is initially in the fully polarized state. The only two collective
pumping operators needed for preparing the many-body singlet from
this initial state are respectively $\sum_{n}\hat{s}_{n}^{-}$ and
$\sum_{n}(-)^{n}\hat{s}_{n}^{+}$. The preparation channel $\mathcal{L}_{P,0}$
is simply realized by the simultaneous pumping with these two operators,
described by the Lindblad form:\begin{equation}
\mathcal{L}_{P,0}\left[\rho\right]=\sum_{j=1,2}\left[\hat{L}_{j}\rho\hat{L}_{j}^{\dagger}-\frac{1}{2}\left(\rho\hat{L}_{j}^{\dagger}\hat{L}_{j}+\hat{L}_{j}^{\dagger}\hat{L}_{j}\rho\right)\right].\label{eq:4-1}\end{equation}
The two Lindblad operators are $\hat{L}_{1}=\sqrt{\Lambda_{h}}\sum_{n}\hat{s}_{n}^{-}$
and $\hat{L}_{2}=\sqrt{\Lambda_{i}}\sum_{n}(-)^{n}\hat{s}_{n}^{+}$,
where $\Lambda_{h}$ and $\Lambda_{i}$ characterize the strength
of the pumping. In the absence of dephasing noises, the population
of many-body singlets $P(J=0) \sim 40\%$ in the steady state when $\Lambda_{h}\gg\Lambda_{i}$
is satisfied. Here, $\Lambda_{i}^{-1}$ characterizes the timescale for
reaching the steady state.

In the presence of inhomogeneous broadening in the qubit resonance,
the steady state will deviate from the target one because of the inhomogeneous
dephasing. We use the population of singlets $P(J=0)$ in the steady
state as the figure of merit. To preserve the collective pumping channel while applying the DD control, in the laboratory
frame, the Lindblad operators shall alternate between two forms in
the even and odd intervals of the DD pulse sequence: \begin{equation}
\hat{L}_{1}\left(t\right)=\begin{cases}
\sqrt{\Lambda_{h}}\sum_{n}\hat{s}_{n}^{-}, & \mbox{even interval},\\
\sqrt{\Lambda_{h}}\sum_{n}\hat{s}_{n}^{+} & \mbox{odd interval.}\end{cases}\label{eq:4-2}\end{equation}
 and \begin{equation}
\hat{L}_{2}\left(t\right)=\begin{cases}
\sqrt{\Lambda_{i}}\sum_{n}(-)^{n}\hat{s}_{n}^{+}, & \mbox{even interval},\\
\sqrt{\Lambda_{i}}\sum_{n}(-)^{n}\hat{s}_{n}^{-}, & \mbox{odd interval.}\end{cases}\label{eq:4-3}\end{equation}

\begin{figure}[ptb]
\begin{centering}
\includegraphics[bb=20 387 549 763,clip,width=3in]{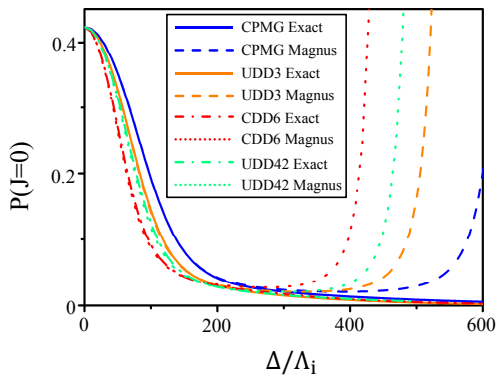}
\par\end{centering}
\caption{(Color online) Convergence of the Magnus expansion. Solid and dot-dashed curves show
the steady state population of many-body singlet $P(J=0)$ obtained
by exactly solving the master equation. Dashed and dotted curves are steady state $P(J=0)$ obtained from the Magnus expansion Eq.~(\ref{eq:2-3}) where we only keep the leading higher order Magnus term (see text). The ensemble
consists of 6 qubits, with inhomogeneously broadened energies $\omega_{i}=\Delta(7-2i)/5$,
$i=1,2,\ldots,6$. The collective pumping rates are chosen as $\Lambda_{h}=100\Lambda_{i}$. The blue, orange, red and green solid lines represent
numerical results when the basic unit of DD control uses the CPMG,
UDD3, CDD6 and UDD42 pulse sequences, respectively. The average pulse
density is chosen as $\bar{n}\equiv N/t_{p}=10^{2.75}\Lambda_{i}$.}
\label{fig:fig5}
\end{figure}

\begin{figure}[ptb]
\begin{centering}
\includegraphics[bb=101 488 479 773,clip,width=3in]{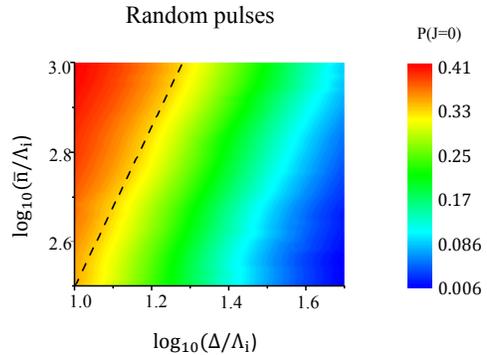}
\par\end{centering}
\caption{(Color online) Steady state population of the many-body singlet $P(J=0)$
as function of the inhomogeneous broadening $\Delta$ and the average
pulse density $\bar{n}=N/t_{p}$ when the DD control uses completely random pulse
sequence. The ensemble consists of 6 qubits, with inhomogeneously
broadened energies $\omega_{i}=\Delta(7-2i)/5$, $i=1,2,\ldots,6$.
The collective pumping rates are chosen as $\Lambda_{h}=100\Lambda_{i}$.}
\label{fig:fig6}
\end{figure}

The numerical simulations are shown in Fig.~3 - 6. In Fig.~\ref{fig:fig3},
we show the state preparation efficiency with the increase of the
inhomogeneous broadening when the preparation process is protected
with various pulse sequences. The steady state population of many-body
singlet $P(J=0)$ is obtained by solving the master equation exactly.
From Fig.~\ref{fig:fig3}(a), apparently, the state preparation can
tolerate larger inhomogeneous broadening when the basic unit of DD
control uses higher order concatenation designs or Uhrig designs when
$t_{p}$ is fixed. Interestingly, nearly identical protection efficiencies
are observed for CDD3 and UDD5, and also for CDD4 and UDD10. We find
the common feature between CDD3 and UDD5, and between CDD4 and UDD10
is that they consist of the same number of $\pi$ pulses. This is consistent
with the analysis in section \ref{inhomoanalysis}, namely, the efficiency
of protection from inhomogeneous dephasing is largely determined by
the density of pulses, rather than the actual design of the sequence.
In Fig.~\ref{fig:fig3}(c), we compare the performance of various
DD controls by fixing the average pulse interval $\bar{\tau}=t_{p}/N$,
and different pulse sequences indeed exhibit nearly identical protection
efficiency.

In Fig.~\ref{fig:fig4}, we examined the protection efficiency as
function of both the inhomogeneous broadening and the average pulse
density $\bar{n}\equiv1/\bar{\tau}$. For the four DD controls compared,
nearly identical protection efficiency is seen. Moreover, in the double
$\log_{10}$ plot, the contour lines are nearly straight lines with slope 1,
showing that the strength of the residue noise scales as $\Delta^{2}\bar{\tau}^{2}$.
This suggests that $\alpha_{3,b}N^{2}\bar{\tau}^{2}\left[\mathcal{L}_{\mathrm{N}},\left[\mathcal{L}_{\mathrm{P},0},\mathcal{L}_{\mathrm{N}}\right]\right]$
is indeed dominating over other higher order Magnus terms and captures
the major effect of the residue noise (c.f. Eq.~(\ref{leading})).

In Fig.~\ref{fig:fig5}, we check the convergence of the Magnus expansion
by comparing the steady state $P(J=0)$ numerically solved from the
exact master equation Eq.~(\ref{eq:2-1}) with that obtained from the
Magnus expansion Eq.~(\ref{eq:2-3}) where we only keep the leading
higher order Magnus term, i.e. by taking $\Omega(t)=t\left(\mathcal{L}_{P,0}+\alpha_{3,b}N^{2}\bar{\tau}^{2}\left[\mathcal{L}_{\mathrm{N}},\left[\mathcal{L}_{\mathrm{P},0},\mathcal{L}_{\mathrm{N}}\right]\right]\right)$.
Excellent convergence is found even when the inhomogeneous noise is
strong enough to diminish the steady state population of the singlets.

Motivated by the finding that the protection efficiency is determined
by the average pulse density and is nearly independent of the pulse
structure, we further tested the possibility of protecting the state
preparation process by a completely random sequence of $\pi$-pulses.
Since the DD control in such case does not have the periodic structure,
the analytical results of Eq.~(\ref{eq:3-4}) and Eq.~(\ref{3-5}) are no
longer applicable. Nevertheless, as shown in Fig.~\ref{fig:fig6}, the state preparation can still be well
protected from inhomogeneous dephasing.
The slope of the contour line is close to 2 in the shown parameter
regime, from which we can tell that the residue noise scales as $\Delta^{2}\bar{\tau}$,
different from the periodic DD controls where
the basic unit uses CPMG, CDD and UDD designs.

Our numerical study shows that sequences of $\pi$ pulses with random arrival times are also efficient in protecting the dissipative state preparation. This means that the protection is insensitive to the errors in the arrival time of the $\pi$ pulses and accurate control on the arrival times is not necessary. The only temporal control needed is that the $\pi$ pulses have to be synchronized with the switches of the pumping operators in the laboratory frame.

\section{DD protection on the dissipative preparation of linear cluster states}

It has also been shown that collective pumping can be used to prepare
linear cluster states in an atomic ensemble embedded in a cavity~\cite{10}. Here we investigate the DD protection of this dissipative preparation scheme.

\begin{figure}[ptb]
\begin{centering}
\includegraphics[bb=32 226 539 786,clip,width=3in]{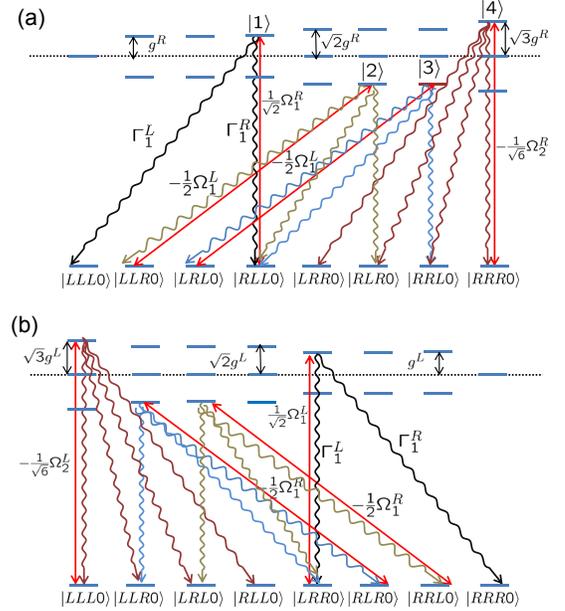}
\par\end{centering}
\caption{(Color online) Level scheme for realizing the stabilizer $S_{3}=\hat{s}_{1}^{z} \hat{s}_{2}^{z} \hat{s}_{3}^{z}$. In the notations such as $|RLL0\rangle$, the first three letters denote the state of qubit 1, 2, 3 respectively and the last integer denotes the number of cavity photons. The four involved states in the excited state manifold are: $\left|1\right\rangle \equiv1/\sqrt{2}\left(\left|RLL1\right\rangle +\left|ELL0\right\rangle \right)$, $\left|2\right\rangle \equiv1/\sqrt{2}\left|RLR1\right\rangle -1/2 (\left|ELR0\right\rangle +\left|RLE0\right\rangle )$, $\left|3\right\rangle \equiv1/\sqrt{2}\left|RRL1\right\rangle -1/2 (\left|ERL0\right\rangle +\left|REL0\right\rangle )$, $\left|4\right\rangle \equiv1/\sqrt{2}\left|RRR1\right\rangle -1/\sqrt{6} (\left|ERR0\right\rangle +\left|RER0\right\rangle +\left|RRE0\right\rangle)$. The wavy lines denote the spontaneous emission. Optical fields resonantly pump the four transitions denoted by the double head solid lines. The eigenstate of stabilizer $S_{3}$ is realized as the steady state of the pumping. (a) and (b) correspond respectively to pumping controls in the laboratory frame before and after a $\pi$ pulse is applied. In (a) the cavity couples to the $\left|R\right\rangle \leftrightarrow \left|E\right\rangle$ transitions with strength $g^R$, while in (b) the cavity couples to the the $\left|L\right\rangle \leftrightarrow \left|E\right\rangle$ transitions with strength $g^L$~\cite{38}. In the toggling frame, the two pumping controls realize the same dissipative channel.}
\label{fig:fig7}
\end{figure}

Key to the preparation of linear cluster states is to realize the stabilizers
through the competition between the optical pumping and the atomic spontaneous emissions. The level scheme for realizing the
stabilizer $S_{3}=\hat{s}_{1}^{z} \hat{s}_{2}^{z} \hat{s}_{3}^{z}$ is shown schematically in Fig.~\ref{fig:fig7}. The qubit is defined in the ground state manifold $\{|L\rangle_j, |R\rangle_j \}$ of a three level atom. Optical fields denoted by the double head solid lines in Fig.~\ref{fig:fig7} couple the two ground states to the common excited state $| E \rangle_j$ with Rabi frequencies $\Omega^L_j$ and $\Omega^R_j$ respectively, where $j$ labels the atoms. All atoms are coupled to a common cavity mode with coupling strength $g$. $\Gamma_{j}^{L}$ and $\Gamma_{j}^{R}$ are the spontaneous emission rate from $| E \rangle_j$ to the two ground states respectively, whose values are usually considered as the same. Those spontaneous emissions are denoted by wavy lines in Fig.~\ref{fig:fig7}. When the driving fields are tuned in resonances with the transitions as indicated in the figure, the eigenstate of stabilizer $S_{3}$ is realized as the steady state of the pumping. Together with single qubit rotations, a complete group of stabilizers for $n-$qubit linear cluster states
can be realized one by one (e.g., $S_{4}^1=\hat{s}_{1}^{x} \hat{s}_{2}^{z}$, $S_{4}^2=\hat{s}_{1}^{z} \hat{s}_{2}^{x} \hat{s}_{3}^{z}$, $S_{4}^3=\hat{s}_{2}^{z} \hat{s}_{3}^{x} \hat{s}_{4}^{z}$ and  $S_{4}^4=\hat{s}_{3}^{z} \hat{s}_{4}^{x}$ for 4-qubit linear cluster states). And thus the fidelity of the linear cluster states can
approach unity when damping rate of atomic spontaneous emission $\Gamma$ is much smaller than the cavity-atom coupling strength $g$. More details of the preparation scheme can be found in Ref.~\cite{10}.

The scheme relies on delicate engineering of detunings, hence the preparation fidelity will be substantially affected by inhomogeneous broadening of the qubit resonances. We consider interlacing DD pulse sequences with the pumping to suppress the effect of this noise. Since the $\pi-$pulse on the qubit simply switches $|L\rangle_{j}$
and $|R\rangle_{j}$ as to preserve the desired form of pumping channel in the toggling frame, one simply needs to switch values for every pair of Rabi frequencies $\Omega_j^L$ and $\Omega_j^R$ and corresponding frequencies of the pump fields in the laboratory frame whenever a $\pi$-pulse is applied in the DD control. The required switching of the controls in the laboratory frame is illustrated in Fig.~\ref{fig:fig7}, where part (a) and (b) correspond respectively to the control before and after a $\pi$-pulse is applied. This is equivalent to using different optical pumping fields in the even and odd intervals of the DD control.

\begin{figure}[ptb]
\begin{centering}
\includegraphics[bb=23 371 560 796,clip,width=3in]{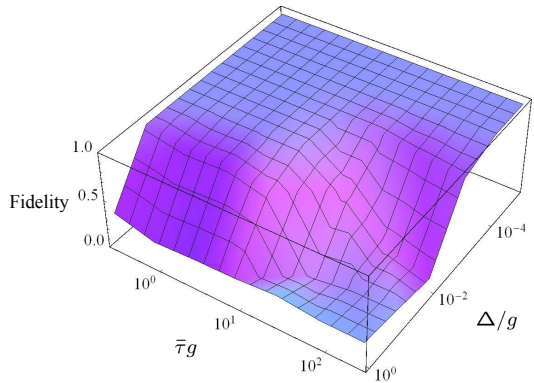}
\par\end{centering}
\caption{(Color online) Fidelity between the steady state and the target
$4$-qubit linear cluster state under the DD protection where the basic unit uses the UDD3 sequence. The inhomogeneously broadened qubit resonances are
$\omega_{i}=\Delta (5-2i)/3$, $i=1,2,3,4$. The parameters for the preparation channel are: $\Omega_{j}^{L}=\Omega_{j}^{R}=0.01g$, $\Gamma_{j}^{L}=\Gamma_{j}^{R}=0.01g$. See text.}
\label{fig:fig8}
\end{figure}

In Fig.~\ref{fig:fig8}, we show numerical simulation on the protection of preparing
a $4$-qubit linear cluster state. The parameters $\Gamma_{j}^{L}=\Gamma_{j}^{R}=0.01g$
and $\Omega_{j}^{L}=\Omega_{j}^{R}=0.01g$. The basic unit of the periodic DD control uses the UDD3 sequence. Indeed, the DD protection is efficient in protecting this dissipative preparation scheme when the inhomogeneous broadening $\Delta \leq 0.1 g$. The fidelity of the steady state with the target one approaches unity when the average pulse interval $\bar{\tau}$ is small.

\section{DD Protection from dynamic noises}

Dynamic fluctuation in the environment can also lead to qubit dephasing. Here we investigate the possibility of DD
protection of the dissipative state preparation in the presence of
dynamic noise, for which we assume a semiclassical form. The
qubit energies are given by $H_{\mathrm{SC}}=\sum_{i}B_{i}\left(t\right)\sigma_{i}^{z}$,
where $B_{i}(t)$ is stochastic and is characterized by the autocorrelation
functions $\left\langle B_{i}\left(t\right)B_{j}\left(t'\right)\right\rangle =G\left(t-t'\right)\delta_{i,j}$.
We further assume the noises are Gaussian, for which we can derive
the Kubo stochastic Liouville equation~\cite{39} in the toggling
frame as \begin{equation}
\dot{\rho}=\mathcal{L}_{\mathrm{P,0}}\left[\rho\right]+\mathcal{L}_{\mathrm{N}}^{\mathrm{eff}}\left[t;\rho\right]\label{eq:6-1}\end{equation}
where the effective noise operator is \begin{equation}
\mathcal{L}_{\mathrm{N}}^{\mathrm{eff}}\left[t;\rho\right]=\sum_{i}\int_{0}^{t}dt_{1}G\left(t-t_{1}\right)f(t)f(t_{1})\mathcal{L}^{i}\left(t-t_{1}\right)\label{eq:6-2}\end{equation}
with $\mathcal{L}^{i}\left(t\right)=\mathcal{L}_{\mathrm{V}}^{i}\mathcal{\widetilde{L}}_{\mathrm{V}}^{i}\left(t\right)$,
where $\mathcal{L}_{\mathrm{V}}^{i}\left[\rho\right]=-i\left[\sigma_{i}^{z},\rho\right]$
and $\mathcal{\widetilde{L}}_{\mathrm{V}}^{i}\left(t\right)=\exp\left(\mathcal{L}_{\mathrm{P,0}}t\right)\mathcal{L}_{\mathrm{V}}^{i}\exp\left(-\mathcal{L}_{\mathrm{P,0}}t\right)$.
$f(t)=(-)^{i}$ for $\tau_{i}<t<\tau_{i+1}$ is the step-like function
corresponding to the DD control.

Following Eq. (\ref{eq:2-4-1}), the first term of the Magnus series
can be written as
\begin{equation}
\Omega_{1}\left(t\right)=t\mathcal{L}_{\mathrm{P,0}}\left[\rho\right]+\int_{0}^{\infty}G\left(\omega\right)\left|F\left(\omega,t\right)\right|d\omega,\label{eq:6-3}
\end{equation}
where $G\left(\omega\right)=1/2\pi\int_{-\infty}^{\infty}G\left(t\right) \exp(-i\omega t)dt$
is the noise spectrum, and $F$ is the filter function corresponding
to the DD control used~\cite{30,40,41,42}
\begin{equation}
\left|F\left(\omega,t\right)\right|=\sum_{i}\int_{0}^{t}\int_{0}^{t_{1}}\frac{dt_{1}dt_{2}}{2\pi}f(t_{1})f(t_{1}-t_{2})\cos\omega t_{2}\mathcal{L}^{i}\left(t_{2}\right).\label{eq:6-4}
\end{equation}

In the case of protecting quantum memory, the preparation channel $\mathcal{L}_{\mathrm{P,0}}$ is absent. Thus $\mathcal{L}^{i}\left(t_{2}\right)=\left(\mathcal{L}_{\mathrm{V}}^{i}\right)^{2}$, which is time-independent and can be taken out of the integral in Eq.~(\ref{eq:6-4}). In such case, concatenated DD or Uhrig DD pulse sequences can generate the filter function that suppresses dynamic noise to arbitrarily high order of pulse intervals~\cite{30,40,41,42}.

However, for the protection of a dissipative state preparation, because of the presence of the nontrivial preparation channel $\mathcal{L}_{\mathrm{P,0}}$, $\mathcal{L}^{i}\left(t_{2}\right)$ can be time-dependent due
to the interference between $\mathcal{L}_{\mathrm{P,0}}$ and $\mathcal{L}_{\mathrm{V}}^{i}$.
The effect of dynamics noises can not be arbitrarily suppressed as in the case of DD protection of quantum memory. In general, $F$ scales as $\bar{\tau}^{2}$, and the dynamical noise can be suppressed to the second-order
of the pulse interval.

\section{Summary}

We investigated the protection of dissipative quantum state preparation processes
against qubit dephasing by interlacing the DD control pulse
sequences with the preparation control. The basic idea is to average out the noise channels over time, while preserving the irreversible dynamics for generating desired entanglement in the steady state. By utilizing
DD control consisting of sequences of short $\pi$-pulses, the dephasing noise can be suppressed to certain order of pulse interval and thus high-fidelity state preparation can be realized when the DD control has sufficiently small pulse interval. With the help of generalized Magnus series, we investigated the order of suppression of noise channels. For inhomogeneous dephasing, the leading noise term in the Magnus expansion is of second order of the pulse interval, and is largely determined by the average pulse interval rather than the types of DD control sequences used. The DD protection efficiency is demonstrated by numerical simulations on two exemplary state preparation schemes for realizing many-body singlets and linear cluster states respectively. DD protection from dynamical noise is discussed for the example of Gaussian noise of a semiclassical description, where the leading noise effect is also of second order of the pulse interval.

\begin{acknowledgements}

WY thanks CQI at IIIS of Tsinghua for hospitality during his visit
through the support by NBRPC under grant 2011CBA00300 (2011CBA00301).
The work was supported by the Research Grant Council of Hong Kong
under Grant No. HKU 706309P and HKU8/CRF/11G. The authors acknowledge
helpful discussion with Hongyi Yu.

\end{acknowledgements}


\begin{thebibliography}{44}


\bibitem{1} S. Clark, A. Peng, M. Gu, and S. Parkins, Phys. Rev.
Lett. \textbf{91}, 177901 (2003).

\bibitem{2} S. Schneider and G. J. Milburn, Phys. Rev. A \textbf{65},
042107 (2002).

\bibitem{3} M. Paternostro, W. Son, and M. S. Kim, Phys. Rev. Lett.
\textbf{92}, 197901 (2004).

\bibitem{4} M. J. Kastoryano, F. Reiter, and A. S. S\o rensen, Phys.
Rev. Lett. \textbf{106}, 090502 (2011).

\bibitem{5} F. Verstraete, M. M. Wolf, and J. I. Cirac, Nat. Phys.
\textbf{5}, 633 (2009).

\bibitem{6} S. Diehl, A. Micheli, A. Kantian, B. Kraus, H. P. B\"{u} chler,
and P. Zoller, Nat. Phys. \textbf{4}, 878 (2008).

\bibitem{7} B. Kraus, H. P. Büchler, S. Diehl, A. Kantian, A. Micheli,
and P. Zoller, Phys. Rev. A \textbf{78}, 042307 (2008).

\bibitem{8} H. Krauter, C. A. Muschik, K. Jensen, W. Wasilewski, J. M. Petersen, J. I. Cirac, and E. S. Polzik, Phys. Rev. Lett. \textbf{107},
080503 (2011).

\bibitem{9} H. Weimer, M. Müller, I. Lesanovsky, P. Zoller and H.
P. Büchler, Nat. Phys. \textbf{6}, 382 (2010).

\bibitem{10} J. Cho, S. Bose, and M. S. Kim, Phys. Rev. Lett. \textbf{106},
020504 (2011).

\bibitem{11} W. Yao, Phys. Rev. B \textbf{83}, 201308(R) (2011).

\bibitem{12} H.-Y. Yu, Y. Luo, and W. Yao, Phys. Rev. A \textbf{84},
032337 (2011).

\bibitem{13} Y. Luo, H.-Y. Yu, and W. Yao, Phys. Rev. B \textbf{85},
155304 (2012).

\bibitem{1b} D. Gottesman and I. L. Chuang, Nature \textbf{402}, 390 (1999).

\bibitem{2b} M. Murao, D. Jonathan, M. B. Plenio, and V. Vedral, Phys. Rev. A \textbf{59}, 156 (1999).

\bibitem{3b} R. Raussendorf and H. J. Briegel, Phys. Rev. Lett. \textbf{86}, 5188 (2001).

\bibitem{4b} H. J. Briegel, D. E. Browne,W. Dür, R. Raussendorf and M. Van den Nest, Nat. Phys. \textbf{5}, 19 (2009).

\bibitem{14} E. L. Hahn, Phys. Rev. \textbf{80}, 580 (1950).

\bibitem{15} M. Mehring, \textit{Principles of High Resolution NMR
in Solids} (Spinger-Verleg, Berlin, 1983), 2nd ed.

\bibitem{16} W.-K. Rhim, A. Pines, and J. S. Waugh, Phys. Rev. Lett.
\textbf{25}, 218 (1970).

\bibitem{17} U. Haeberlen, \textit{High resolution NMR in solids:
selective averaging} (Academic Press, New York, 1976).

\bibitem{18} L. Viola and S. Lloyd, Phys. Rev. A \textbf{58}, 2733
(1998).

\bibitem{19} M. Ban, J. Mod. Opt. \textbf{45}, 2315 (1998).

\bibitem{20} P. Zanardi, Phys. Lett. A \textbf{258}, 77 (1999).

\bibitem{21} L. Viola, E. Knill, and S. Lloyd, Phys. Rev. Lett. \textbf{82},
2417 (1999).

\bibitem{36} D. A. Lidar, Phys. Rev. Lett. \textbf{100}, 160506 (2008).

\bibitem{gate1} C. Barthel, J. Medford, C. M. Marcus, M. P. Hanson, and A. C. Gossard, Phys. Rev. Lett. \textbf{105}, 266808 (2010).

\bibitem{gate} T. van der Sar, Z. H. Wang,	M. S. Blok, H. Bernien, T. H. Taminiau, D. M. Toyli, D. A. Lidar, D. D. Awschalom, R. Hanson, and V. V. Dobrovitski, Nature \textbf{484}, 82 (2012).

\bibitem{gate2} J. R. West, D. A. Lidar, B. H. Fong, and M. F. Gyure, Phys. Rev. Lett. \textbf{105}, 230503 (2010).

\bibitem{gate3} X. K. Xu, Z. X. Wang, C. K. Duan, P. Huang, P. F. Wang, Y. Wang, N. Y. Xu, X. Kong, F. Z. Shi, X. Rong, and J. F. Du, Phys. Rev. Lett. \textbf{109}, 070502 (2012).

\bibitem{22} H. Carr and E. M. Purcell, Phys. Rev. \textbf{94}, 630
(1954).

\bibitem{23} S. Meiboom and D. Gill, Rev. Sci. Instrum. \textbf{29},
688 (1958).

\bibitem{24} K. Khodjasteh and D. A. Lidar, Phys. Rev. Lett. \textbf{95},
180501 (2005).

\bibitem{25} K. Khodjasteh and D. A. Lidar, Phys. Rev. A \textbf{75},
062310 (2007).

\bibitem{26} W. Yao, R. B. Liu, and L. J. Sham, Phys. Rev. Lett.
\textbf{98}, 077602 (2007).

\bibitem{27} R. B. Liu, W. Yao, and L. J. Sham, New J. Phys. \textbf{9},
226 (2007).

\bibitem{28} W. M. Witzel and S. Das Sarma, Phys. Rev. B \textbf{76},
241303(R) (2007).

\bibitem{29} W. X. Zhang, V. V. Dobrovitski, L. F. Santos, L. Viola,
and B. N. Harmon, Phys. Rev. B \textbf{75}, 201302(R) (2007).

\bibitem{30} G. S. Uhrig, Phys. Rev. Lett. \textbf{98}, 100504 (2007).

\bibitem{31} B. Lee, W. M. Witzel, and S. Das Sarma, Phys. Rev. Lett.
\textbf{100}, 160505 (2008).

\bibitem{32} G. S. Uhrig, New J. Phys. \textbf{10}, 083024 (2008).

\bibitem{33} W. Yang and R.-B. Liu, Phys. Rev. Lett. \textbf{101},
180403 (2008).

\bibitem{34} L. Viola and E. Knill, Phys. Rev. Lett. \textbf{94},
060502 (2005).

\bibitem{35} L. F. Santos and L. Viola, Phys. Rev. Lett. \textbf{97},
150501 (2006).

\bibitem{37} S. Blanes, F. Casas, J. A. Oteo, and J. Ros, Phys. Rep. \textbf{470}, 151 (2009).

\bibitem{38} Such tunable cavity-atom couplings are required by the original proposal in Ref.~\cite{10}.

\bibitem{39} R. Kubo, J. Math. Phys. \textbf{4}, 174 (1963).

\bibitem{40} A. G. Kofman and G. Kurizki, Phys. Rev. Lett. \textbf{93}, 130406 (2004).

\bibitem{41} H. Uys, M. J. Biercuk, and J. J. Bollinger, Phys. Rev. Lett. \textbf{103}, 040501 (2009).

\bibitem{42} \L. Cywi\'{n}ski, R. M. Lutchyn, C. P. Nave, and S. Das Sarma, Phys. Rev. B \textbf{77}, 174509 (2008).

\end{thebibliography}
\end{document}